\documentstyle[aps,psfig]{revtex}

\begin{document}
\twocolumn[\hsize\textwidth\columnwidth\hsize\csname
@twocolumnfalse\endcsname

\title{Unusual Coarsening During Phase Separation in Polymer Systems}

\author{C. B. Muratov}

\address{Courant Institute of Mathematical Sciences, New York
  University, \\ 251 Mercer St., New York, NY 10012}

\date{\today}

\draft

\maketitle

\begin{abstract}

We introduce a kinetic model of coarsening of transient polymer
networks during the intermediate stages of phase separation in polymer
systems. The model explicitly takes into account the effect of the
connectivity of polymer chains. We show that during the intermediate
stages the size of the droplet grows according to an unusual
coarsening law $R \sim t^{1/(d + 3)}$. When the network structure
breaks up, the coarsening law may cross over to $R \sim \ln t^{-1}$,
until only at very late stages the Lifshitz-Slyozov coarsening law $R
\sim t^{1/3}$ is recovered.

\end{abstract}

\pacs{PACS Number(s): 64.75.+g, 61.25.Hq, 61.41.+e, 05.70.Fh}

\bibliographystyle{prsty}

\vskip2pc]

Kinetics of phase separation in polymer systems is an important
problem both from the scientific and technological point of
view. Phase separation in such systems is driven by the difference in
the chemical potential of the species \cite{flory,degennes,grosberg},
so at long times the kinetics of this process should fall in the same
dynamic universality class as simple fluids \cite{gunton83}. This is
indeed observed in the experiments on phase separation in polymer
solutions and blends (see, for example,
\cite{hashimoto86,bates89,haas97}). Nevertheless, because of the large
size of polymer molecules the universal kinetics of phase separation
may not be seen until very late times. Recent experiments
\cite{tanaka92,tanaka93} and molecular dynamics simulations
\cite{bhattacharya98} showed that in the intermediate stages of phase
separation interconnected network structures may be observed. In other
words, in the intermediate stages the system may become a {\em
transient gel} \cite{tanaka97}, so its connectivity should play
crucial role in the kinetics of phase separation.

In this Letter, we introduce a kinetic model of coarsening of polymer
networks during the intermediate stages of phase separation. For
definiteness, we will consider phase separation in concentrated
polymer solutions undergoing a deep quench into the unstable state. We
will show that the connectivity of the droplets of polymer-rich phase
entirely changes the transport mechanism which determines the
coarsening of droplets, resulting in an unusual coarsening law: $R
\sim t^{1/(d + 3)}$. This coarsening will arrest when the distance
between the droplets reaches the value of the order of the radius of
gyration $R_g$ of the polymer chain before the quench, when the
network structure will break up. After that, a much slower coarsening
mechanism may switch on resulting in the coarsening law $R \sim \ln
t^{-1}$. Only when the distance between droplets becomes of the order
of the length of the fully stretched polymer chain, the coarsening
will cross over to the conventional Lifshitz-Slyozov mechanism with $R
\sim t^{1/3}$ \cite{lifshitz61}.

\begin{figure}[htbp]
\centerline{\psfig{figure=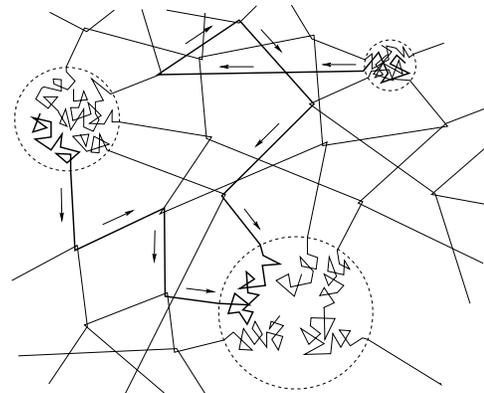,width=2.5in,angle=-90}}
\caption{Transient polymer network. Thick line shows a test chain. The
arrows indicate the motion of the test chain.}
\label{f1}
\end{figure}

When a polymer solution is quenched deep into the unstable region,
spinodal decomposition occurs. After a relatively short transient
period the system segregates into polymer-rich droplets of size of
order $a$, where $a$ is the statistical segment length. Due to
repulsive interaction between the polymer and the solvent polymer-rich
droplets will exert attractive force of order $T / a$ on the chains
that cross their interfaces, pulling them into the droplets. Here two
qualitatively different situations are possible. If the chains are not
very long, the initial transient time will be enough for a droplet to
pull in all the chains that come into contact with it, so after the
transient the morphology of the droplets will be disconnected. In
contrast, if the length of the chains is large enough, during the
initial transient time the chains will be pulled into droplets until
they will become fully stretched between the points of entanglement
(Fig. \ref{f1}). The reason for the latter is that when the chains are
sufficiently long (or, equivalently, when the distance between the
droplets is smaller than the radius of gyration $R_g$ of the chain
before the quench), a single chain may initially become a part of
several droplets at the same time. Therefore, after the transient
period, the morphology of the droplets will be connected. In a
stretched state the conformation of a chain connecting a pair of
droplets will be Gaussian with the effective segment length $a
N_e^{1/2}$, where $N_e$ is the average number of segments between the
entanglement points before the quench (Fig. \ref{f1}).

After the network morphology is formed, the motion of the chains will
significantly slow down since on average the force acting at the point
of attachment of the chain at one droplet will be compensated by the
opposite force from another droplet, thus stretching the chain. The
total force, however, will not be exactly equal to zero since the
droplets between which a piece of the chain is stretched will not
generally have the same radii. The latter will result in the
difference of the surface pressures leading to the net force of order
$T / R$, where $R$ is the characteristic radius of a droplet, acting
on the stretched piece of the chain (Fig. \ref{f1}). So, a piece of
the chain connecting a pair of droplets will slowly creep along itself
from the droplet with smaller radius to the droplet with greater
radius (Fig. \ref{f1}) with the speed of order $T \mu a^2 N_e^{1/2} /
(R L^2)$, where $\mu$ is the mobility of an individual segment and $L$
is the distance between the connected droplets. In estimating the
speed of the chain we assumed that the friction force is proportional
to the number of segments $L^2 /(N_e^{1/2} a^2)$ in the chain
piece. This creeping motion should result in the transport of segments
from the smaller droplets to the larger ones and, therefore, lead to
their coarsening. Note that the connectivity of the chains results in
the {\em nonlocal} mass transport, since for the latter the creep of
the chain is equivalent to taking monomers from one droplet and
putting them instantaneously into the other.

The mass current into a droplet is proportional to the average speed
of the chains into that droplet and the number of chains that come
into it. A peculiar feature of the considered transport mechanism is
the fact that as long as the average droplet size is smaller than the
distance $a N_e^{1/2}$ between entanglements, the number of chains
that come out of a given droplet is independent of time (see
Fig. \ref{f1}). So, the current into the droplet is simply
proportional to the average speed of the chains. This is in contrast
with the conventional Lifshitz-Slyozov mechanism in which the mass
current into the droplet is proportional to the surface area of the
droplet \cite{lifshitz61}. Writing the conservation of mass for a
droplet of size $R$, using the expression for the velocity of a
stretched portion of a chain and taking into account the conservation
of volume fraction of the polymer-rich phase, we obtain that on
average $R^{d-1} d R / dt \sim 1/R^3$. This suggests that the
considered mechanism will result in the average droplet size growing
as $R \sim t^{1/(d + 3)}$ and self-similar growth of the droplet
morphology.

Let us demonstrate that this self-similar growth should actually take
place for low polymer volume fraction when the system contains many
droplets of different sizes. For simplicity, let us consider
three-dimensional system. Then the droplets can be assumed to be
spherical and uniformly distributed over space. Let us introduce the
distribution function $f(R, t)$ normalized in such a way that the
integral $\int f dR$ gives the number of droplets per unit
volume. Since after the initial formation of the network no droplets
will be further created, the function $f$ must satisfy Liouville
equation
\begin{equation} \label{f}
{\partial f \over \partial t} = - {\partial \over \partial R} ( m v
R^{-2} f),
\end{equation}
where $m$ is the number of chains coming out of a droplet and $v$ is
the average speed of the chains into the droplet. As was already
pointed out, on the time scale of the considered coarsening process
the number $m$ of chains coming out of a droplet is
time-independent. At the beginning of coarsening all droplets have
roughly the same size of order $a$, so on average the number $m$ will
be the same for all droplets. Therefore, in the following we will
treat $m$ as a constant of order 1. The average velocity $v$ can be
written as
\begin{equation} \label{v}
v = \int dP(R', L) {1 \over L^2} \left( {1 \over R'} - {1
\over R} \right).
\end{equation}
Here~and in Eq. (\ref{f}) an appropriate non-dimensionalization has
been performed. In Eq. (\ref{v}) the term $dP(R', L)$ is the
probability that the droplet of radius $R$ is connected with the
droplet with radius $R'$ at distance $L$, the term $1/L^2$ comes from
the dependence of the mobility of the stretched portion of the chain
on its length, and the term $(1/R' - 1/R)$ is the driving force.

In the following we will assume that the chains connect different
droplets at random. Let us measure distances relative to the droplet
of radius $R$. Since the chains are highly stretched, the probability
$dP$ is essentially the probability that a random walk of step $a
N_e^{1/2}$ starting from the origin terminates at a droplet of size
$R'$ at distance $L$ while not crossing any other droplets on its
way. This probability is the product of two parts: $dP(R', L) =
dP_1(R', L) P_2(L)$, where $dP_1(R', L)$ is the probability that a
droplet of size $R'$ at distance $L$ is connected to the origin
disregarding the effect of other droplets, and $P_2$ is the
probability that the chain connecting these droplets did not cross any
other droplets. Since by assumption the same number of chains comes
out of all droplets regardless of their size, the probability $dP_1$
can be written as
\begin{equation} \label{p1}
dP_1(R', L) = C L^2 f(R') dR' dL,
\end{equation}
where $C$ is the normalization constant. It is not difficult to show
that for a random walk in the presence of small spherical obstacles
with the distribution function $f$
\begin{equation} \label{p2}
P_2(L) = e^{-\kappa L^2}, ~~~~~~\kappa = 4 \pi \int R f dR.
\end{equation}
The constant $C$ is determined by the condition $\int dP = 1$. This
gives us
\begin{equation} \label{C}
C = {4 \kappa^{3/2} \over \sqrt{\pi} \int f dR}.
\end{equation}
Note that $\kappa$ and $C$ are determined by the distribution $f$ and
therefore are functions of time.

Writing the probability $dP$ in the above form essentially constitutes
a self-similarity assumption about the connectivity of the network as
a function of time. This should be a good assumption since as smaller
droplets collapse during coarsening, the chains that came out of them
simply reconnect between other droplets.

Having now calculated $dP$, we can substitute it to Eq. (\ref{v}) and
integrate over $L$. As a result, we can write the expression for $v$
(absorbing $m$ into it) in the form
\begin{equation} \label{vv}
v = \Lambda {\int R f dR \over \int f dR} \int dR' f(R') \left(
{1 \over R'} - {1 \over R} \right),
\end{equation}
where $\Lambda = 6 m \phi$, $\phi$ is the volume fraction of the
polymer-rich phase and the normalization of $f$ was changed to $\int
R^3 f dR = 1$. The constant $\Lambda$ determines the rate of the
coarsening process. In the following we will absorb it into the
definition of $t$. Note that according to Eq. (\ref{vv}) we have $\int
v f dR = 0$, meaning the conservation of the overall mass of droplets
(assuming that the volume fraction of the chains outside the droplets
is negligible).

Equations (\ref{f}) and (\ref{vv}) are of the same type as those
studied by Lifshitz and Slyozov in the theory of coarsening by
evaporation-condensation mechanism \cite{lifshitz61} and can be
treated in the same way. Let us introduce a similarity ansatz $f(R, t)
= t^{-2/3} F \left( R t^{-1/6} \right)$ into Eqs. (\ref{f}) and
(\ref{vv}). Introducing $x = R t^{-1/6}$, we obtain an equation
\begin{equation} \label{self}
4 F + x F' = 6 a_1 {d \over dx} \left[ F \left( {a_2} x^{-2}
- x^{-3} \right) \right],
\end{equation}
where the constants $a_1$ and $a_2$ are given by
\begin{equation} \label{ab}
a_1 = \int_0^\infty x F dx, ~~~~a_2 = {\int_0^\infty x^{-1} F dx
\over \int_0^\infty F dx},
\end{equation}
and the prime denotes differentiation with respect to $x$.  The
function $F$ is normalized according to
\begin{equation} \label{norm}
\int_0^\infty x^3 F dx = 1.
\end{equation}
Note that from the condition of the conservation of mass $\int v f dR
= 0$ follows that the second of Eq. (\ref{ab}) is identically
satisfied for any solution of Eq. (\ref{self}).

\begin{figure}[htbp]
\centerline{\psfig{figure=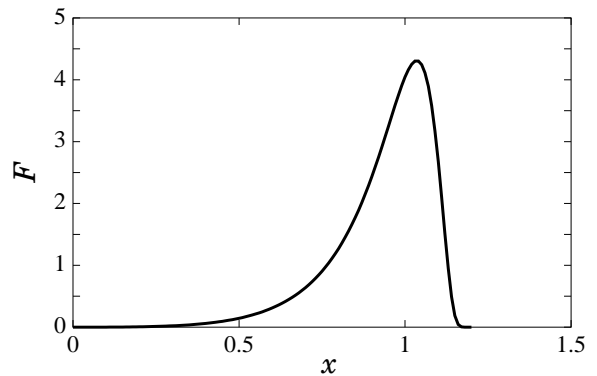,width=3in}}
\caption{The distribution $F(x)$ for $d = 3$.}
\label{f2}
\end{figure}

Equation (\ref{self}) can be straightforwardly integrated 
\begin{equation} \label{sol}
F = A e^{ - \int_1^x {4 y^4 + 12 a_1 a_2 y - 18 a_1 \over y^5 - 6 a_1
a_2 y^2 + 6 a_1 y} dy },
\end{equation}
where the constant $A$ is determined via Eq. (\ref{norm}). As in the
Lifshitz-Slyozov theory \cite{lifshitz61}, the solution that is
selected at long times from the extended initial distribution should
vanish alongside with all its derivatives at some $x = x^*$. This is
only possible if the integrand of Eq. (\ref{sol}) has a double pole at
this value of $x$. The analysis shows that this situation is realized
if for a given value of $a_2$ we have $x^* = 4 / (3 a_2)$ and $a_1 =
2^7 /( 3^4 a_2^4)$. Then, the constant $a_1$ is calculated so that
the first of Eq. (\ref{ab}) is satisfied. We have done this
calculation numerically and found that $a_2 = 1.1027$ and $A =
4.0446$. The solution for $F$ in this case is shown in
Fig. \ref{f2}. From this solution one finds that the average radius of
the droplet grows as $\langle R \rangle = 0.94 t^{1/6}$, while the
average droplet density decays as $\langle n \rangle = 1.14
t^{-1/2}$. Note that the polydispersity of the droplets during
coarsening is rather small, we have $(\langle R^2 \rangle - \langle R
\rangle^2) / \langle R \rangle^2 = 0.023$ in $d = 3$.

Observe that in his pioneering paper on phase separation in polymer
systems de Gennes proposed the coarsening law $R \sim t^{1/5}$ for $d
= 3$ in the early stages of coarsening \cite{degennes80}. In
\cite{degennes80} the transport is due to diffusion of chains between
the droplets with Rouse mobility of the chains \cite{doi} which is
taken to depend on $R$. However, it is clear that Rouse diffusion of
the chains will be suppressed when the distance between the droplets
becomes comparable to the distance between entanglements soon after
the onset of phase separation. After that the system becomes a
transient gel, so the coarsening mechanism should change to the one
studied by us. Note that the importance of highly stretched chains for
phase separation kinetics was pointed out already in
\cite{degennes80}.

The coarsening law $R \sim t^{1/6}$ obtained by us for three
dimensions is in very good agreement with the experimental results of
Tanaka who finds that $R \sim t^{0.15}$ for a phase separating
polymers solution in the network state. Also, the coarsening law $R
\sim t^{0.23}$ observed in experiments on coarsening of thin diblock
copolymer films \cite{coulon93,bassereau93} is in good agreement with
our result for two dimensions.

Note that in the analysis of coarsening we assumed that the number of
chains coming out of a droplet is independent of time. This is a good
assumption as long as the size of the droplet is smaller than the
distance $a N_e^{1/2}$ between the entanglement points. At the onset
of the considered coarsening process the distance between the droplets
is of order $a N_e^{1/2}$, so the radius of a droplet will be much
smaller than that in the case of low polymer volume fraction. In fact,
the smallness of the polymer volume fraction is the necessary
condition for the considered coarsening mechanism to be
realized. Note, however, that before the size of a droplet becomes
comparable to $a N_e^{1/2}$, the distance between the droplets may
reach the value of the order of the radius of gyration $R_g$ of the
chains before the quench, so the network will break up. The latter is
observed in the late stages of coarsening in the experiments of Tanaka
\cite{tanaka93}.

After the breakup of the network one would expect that the
conventional Lifshitz-Slyozov transport mechanism turns on resulting
in $R \sim t^{1/3}$ growth law. Note, however, that the rate constant
for such a mechanism will be extremely small for deep quenches
\cite{degennes80} (for recent developments, see also
\cite{otto97}). This is due to the fact that Lifshitz-Slyozov
mechanism would require an evaporation of an entire chain from a
droplet, which then diffuses through the matrix to another
droplet. For deep quenches the free energy cost of such a fluctuation
will be of order $T N$, where $N$ is the polymerization index, so the
rate at which such an event would occur will be proportional to
$e^{-N}$ which rapidly becomes very small as $N$ increases. Note,
however, that this is not the only possibility for a transport of
chains if the droplets are close enough. Instead of the whole chain
coming out of a droplet, a {\em portion} of a chain containing $Z$
segments may come out. Furthermore, this portion of a chain can be
highly stretched, so the free energy cost of such a fluctuation will
be of order $T Z + T L^2/(a^2 Z)$, where $a$ is the statistical
segment length and $L$ is the distance the chain is stretched. For a
given $L$ this expression has a minimum at $Z \sim L/a$, so the free
energy cost of having a chain piece sticking out a distance $L$ from
the droplet is of order $T L / a \ll TN$ for $L \ll N a$. On the other
hand, if the distance between the droplets is of order $L$, there is a
good chance that such a protruding chain piece will come in contact
with another droplet and stick to it. If this happens, after a
relatively short time the chain will become stretched, and the
transport mechanism discussed above will switch on (Fig. \ref{f1})
leading to a transfer of the entire chain from the droplet with
smaller radius to the one with the greater radius. The rate of such a
process will be proportional to $e^{-L/a}$, so one would expect this
to lead to coarsening of the form $L \sim \ln t^{-1}$. Of course, when
the distance between the droplets becomes comparable to $Na$, such a
process will no longer be feasible, so the conventional
Lifshitz-Slyozov mechanism with the rate proportional to $e^{-N}$ will
be realized.

In determining the dominant coarsening mechanism in a real system, one
should also consider the coalescence mechanism \cite{binder77}. It is
clear that while the system is in the network state this mechanism
cannot work since the stretched chains prevent the droplets from
executing Brownian motion. On the other hand, when the network breaks
up, the mechanism discussed above quickly becomes very slow, so in
polymer solutions, where the mobility of the droplets will be high,
coalescence should become the dominant transport mechanism. This
conclusion is supported by recent experiments on phase separation of
relatively low molecular weight polymer blends \cite{white95}. In
contrast, in polymer melts the mobility of the droplets will be
significantly smaller, so one would expect the mechanism discussed in
the preceding paragraph to be the dominant one. Notice that because
the latter is very slow, in experiment this can be seen as an arrest
and intermittency of the droplet growth. This is indeed observed in
the experiments on polymer melts \cite{takeno97,takeno98}

In conclusion, we have demonstrated that polymer chain connectivity
may have significant effect on the kinetics of phase separation in
polymer systems resulting in the unusual coarsening laws in the
intermediate stages.

This work was supported by AFOSR grant 5-2274-668-103. 

\bibliography{../main}

\narrowtext

\end{document}